\documentclass[manuscript]{aastex}
\usepackage{pdfsync}
\usepackage{multirow}
\usepackage{booktabs}
\usepackage{hyperref}

\shortauthors{Rollett et al.}
\shorttitle{Combined Multipoint Study of an Asymmetric ICME}

\begin{document}

\title{Combined Multipoint Remote and In Situ Observations of the Asymmetric Evolution of a Fast Solar Coronal Mass Ejection}

\author{T. Rollett, C. M\"ostl}
\affil{IGAM-Kanzelh\"ohe Observatory, Institute of Physics, University of Graz, 8010 Graz, Austria}
\affil{Space Research Institute, Austrian Academy of Sciences, 8042 Graz, Austria}
\email{tanja.rollett@gmx.at}

\author{M. Temmer}
\affil{IGAM-Kanzelh\"ohe Observatory, Institute of Physics, University of Graz, 8010 Graz, Austria}

\author{R.A. Frahm}
\affil{Southwest Research Institute, 6220 Culebra Road, San Antonio, TX 78238, USA}

\author{J.A. Davies}
\affil{RAL Space, Rutherford Appleton Laboratory, Harwell Oxford, OX11 0QX, UK}

\author{A.M. Veronig}
\affil{IGAM-Kanzelh\"ohe Observatory, Institute of Physics, University of Graz, 8010 Graz, Austria}

\author{B. Vr{\v s}nak}
\affil{Hvar Observatory, Faculty of Geodesy, University of Zagreb, 1000 Zagreb, Croatia}

\author{U.V. Amerstorfer}
\affil{IGAM-Kanzelh\"ohe Observatory, Institute of Physics, University of Graz, 8010 Graz, Austria}

\author{C.J.\ Farrugia}
\affil{Space Science Center and Department of Physics, University of New Hampshire, Durham, NH 03824, USA}

\author{T.\ {\v Z}ic}
\affil{Hvar Observatory, Faculty of Geodesy, University of Zagreb, 1000 Zagreb, Croatia}

\and

\author{T.L.\ Zhang}
\affil{Space Research Institute, Austrian Academy of Sciences, 8042 Graz, Austria}




\begin{abstract}

We present an analysis of the fast coronal mass ejection (CME) of 2012 March 7, which was imaged by both \textit{STEREO} spacecraft and observed in situ by \textit{MESSENGER}, \textit{Venus Express}, \textit{Wind} and \textit{Mars Express}. Based on detected arrivals at four  different positions in interplanetary space, it was possible to strongly constrain the kinematics and the shape of the ejection. Using the white-light heliospheric imagery from \textit{STEREO-A} and \textit{B}, we derived two different kinematical profiles for the CME by applying the novel constrained self-similar expansion method. In addition, we used a drag-based model to investigate the influence of the ambient solar wind on the CME's propagation. We found that two preceding CMEs heading in different directions disturbed the overall shape of the CME and influenced its propagation behavior. While the Venus-directed segment underwent a gradual deceleration (from $\sim 2700$ km s$^{-1}$ at 15 $R_\odot$ to $\sim 1500$ km s$^{-1}$ at 154 $R_{\odot}$), the Earth-directed part showed an abrupt retardation below 35 $R_{\odot}$ (from $\sim 1700$ to $\sim 900$ km s$^{-1}$). After that, it was propagating with a quasi-constant speed in the wake of a preceding event. Our results highlight the importance of studies concerning the unequal evolution of CMEs. Forecasting can only be improved if conditions in the solar wind are properly taken into account and if attention is also paid to large events preceding the one being studied.

\end{abstract}





%

\section{Introduction}

Coronal mass ejections (CMEs) are impulsive outbursts from the solar corona, carrying a vast amount of magnetized plasma. The plasma signatures of CMEs can be observed in white-light, besides the LASCO/\textit{SOHO} coronagraphs, from two sides using the coronagraphs and heliospheric imagers \citep[HI;][]{eyl09} aboard the \textit{Solar TErestrial RElations Observatory} \citep[\textit{STEREO};][]{kai08}. In situ observations of CMEs are called interplanetary CMEs (ICMEs) and can be detected by a variety of spacecraft at different locations in the inner heliosphere. We use the term ICME, as defined in \cite{rou11}, starting with the shock signature followed by the sheath and the magnetic structure of the CME. Note that some other authors use the term ICME for the ejecta only \cite[e.g.][]{ric10}. CMEs/ICMEs are the main drivers of space weather disturbances and therefore investigating their kinematics and dynamics has attracted plenty of attention over the last years. The \textit{STEREO} mission has given rise to the development of a number of methods to analyze and forecast a CME's direction, speed and arrival time at Earth and other locations in the heliosphere.

Fitting methods, using single-spacecraft \textit{STEREO}/HI observations, are based on the assumptions of constant propagation speed and fixed direction ($\phi$), and assume a geometrical form for the CME. The self-similar expansion geometry \citep[SSE;][]{lug10b,dav12,moedav13} describes the front of a CME as a circle with any fixed angular width. In the two extreme cases of the SSE geometry, the fixed-$\phi$ geometry \citep[F$\phi$;][]{she99,rou08} assumes a negligible extent of the front and the harmonic mean geometry \citep[HM;][]{lug10,moe11} describes a CME as a circle with a width of $180^\circ$, always attached to Sun-center. If more than single-spacecraft observations are available, these geometric methods can also be used to derive a kinematical profile. Using stereoscopic \textit{STEREO}/HI observations, triangulation of the CME front provides information on variations in propagation direction and speed \citep[][]{liu10a,liu10b,lug10b,dav13}.

Another methodology to derive kinematical profiles if only single-spacecraft white-light data are available, is through exploiting the connection with in situ data \cite[][]{moe09a,moe10}. This approach is called the constrained harmonic mean method \cite[CHM;][]{rol12,rol13} and is adapted in this letter to the more generalized SSE geometry.

In this study, we investigate a CME/ICME launched on 2012 March 7, which was already studied using stereoscopic methods by  \cite{liu13} and \cite{dav13}. Due to the favorable locations of Mercury \citep[\textit{MESSENGER};][]{sol01}, Venus \citep[\textit{Venus Express (VEX)};][]{zha06}, Earth \citep[\textit{Wind};][]{lep95}, and Mars \citep[\textit{Mars Express (MEX);}][]{bar04}, it is possible to detect the ICME arrival at four different times and locations in interplanetary space (spacecraft configurations see Figure \ref{fig4}).
In situ measurements at locations other than Earth were not considered in previous studies of this event, although \citet{liu13} indicated that the event impacted Mars.
Here, we connect the arrival times of the CME at these spacecraft to the white-light imagery of \textit{STEREO-A} and \textit{B} by applying the novel constrained SSE (CSSE) method. We obtain the global shape and propagation of the CME with a circular model for its front, constrained by remote white-light images and by in situ observations at various locations. We also take into account the differing solar wind conditions by applying the drag-based model \citep[DBM;][]{vrs12}. Following the approach in \cite{tem12}, we calculate and compare the drag-parameters of two segments of the same CME and relate the results to the solar wind conditions along the two tracks.

\section{Observations}

The event under study occurred within a series of other CMEs, which could have influenced its propagation behavior. In the following, all four CMEs are described in chronological order of their launch and listed in Table \ref{tbl-1}, the CME under study denoted as CME3.

CME1, which entered the field of view (FoV) of COR1-B on 2012 March 5 02:46 UT and of COR1-A at 03:10 UT, was detected  by \textit{MESSENGER} at $\sim$ 12:30 UT and by \textit{VEX} at 08:30 UT on the next day. Using the detected arrival times and assuming a constant propagation speed between Mercury and Venus, we estimate the CME speed to be $\sim$ 840 km s$^{-1}$. CME2 was first observed by COR1-B on 2012 March 6 at 02:36 UT and by COR1-A at 03:25 UT. It was directed toward Earth with a mean speed of $\sim 700$ km s$^{-1}$ during its propagation phase within the HI-FoV, derived by applying the stereoscopic SSE \textbf{(SSSE)} method.

\begin{figure}
\epsscale{0.6}
\plotone{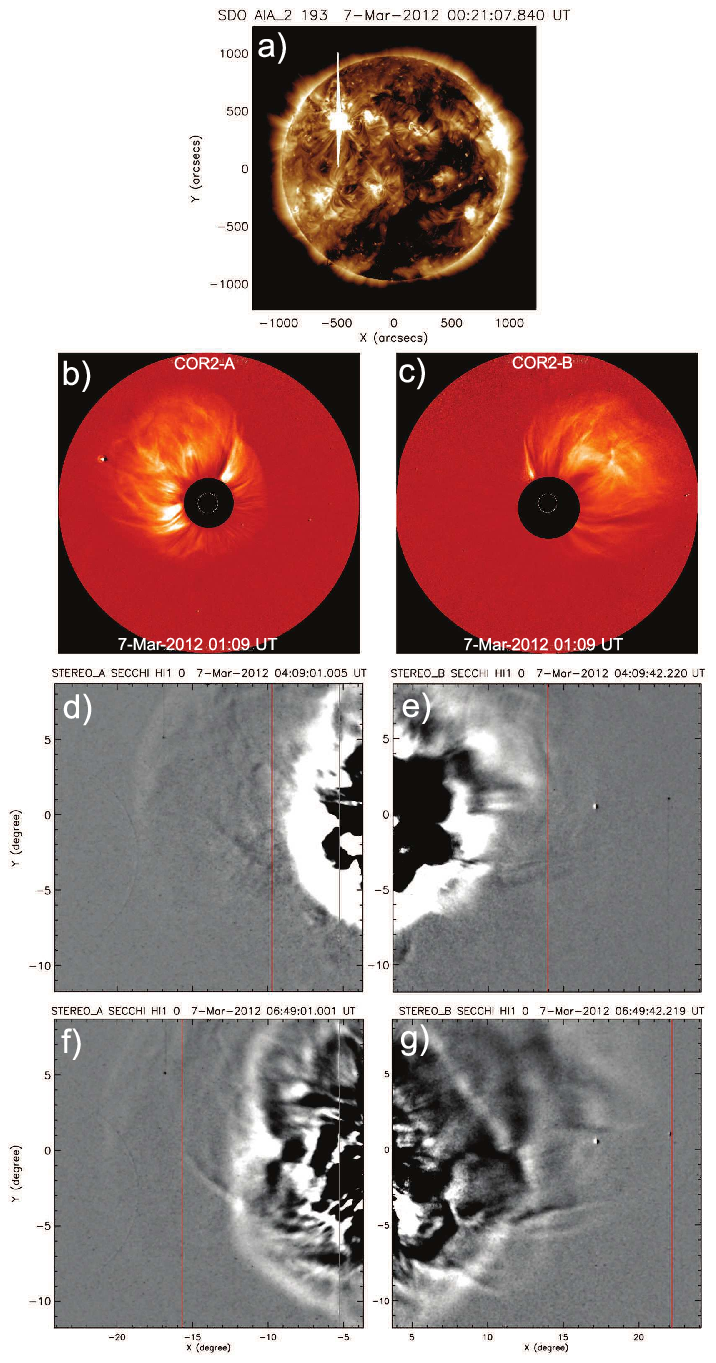}
\caption{Remote observations from three vantage points. a) \textit{SDO}/AIA 193 $\mbox{\AA}$ image at the time of the flare and the CME onset. b) COR2-A (left) and B (right) images. d) and f) HI1-A and e) and g) HI1-B observations of the CME under study (CME3). The red lines mark the position of the elongation measurement. In the HI1-A images, the preceding CME (CME2) is also visible.\label{fig1}}
\end{figure}

For CME3 we use the advantage of three viewing points for remote-sensing observations and in situ measurements from four spacecraft at different positions. Remote observations were performed by \textit{SDO}/AIA \citep[][]{lem12} and \textit{STEREO}/SECCHI \citep[][]{how08}. Using obervations of HI1 and HI2, we were able to follow the CME as it travels from the Sun to Venus (observed within \textit{STEREO-B}/HI) and Earth (observed within \textit{STEREO-A}/HI).
CME3 was associated with a GOES X5.4 flare on 2012 March 7 00:02 UT at N17E16 and a large-scale EUV wave. Figure \ref{fig1}a shows an \textit{SDO}/AIA 193 $\mbox{\AA}$ image from 2012 March 7 at 00:21 UT. CME3 was first observed by COR1-B on 2012 March 7 00:16 UT and by COR1-A at 00:25, entering the FoVs of COR2-B and COR2-A at 00:39 UT. Figures \ref{fig1}b and \ref{fig1}c show images from COR2-A and COR2-B at 01:09 UT. At 00:49 UT and 01:29 UT CME3 reached the FoVs of HI1-B and HI1-A, respectively. Figures \ref{fig1}d--g show CME3 in the HI1-A (left) and HI1-B (right) FoVs at two different times. The red vertical lines overlaid on the HI1 images indicate where the measurements are taken. At 06:10 UT and 16:09 UT CME3 reached the FoVs of HI2-B and HI2-A, respectively.

\begin{figure}
\epsscale{0.6}
\plotone{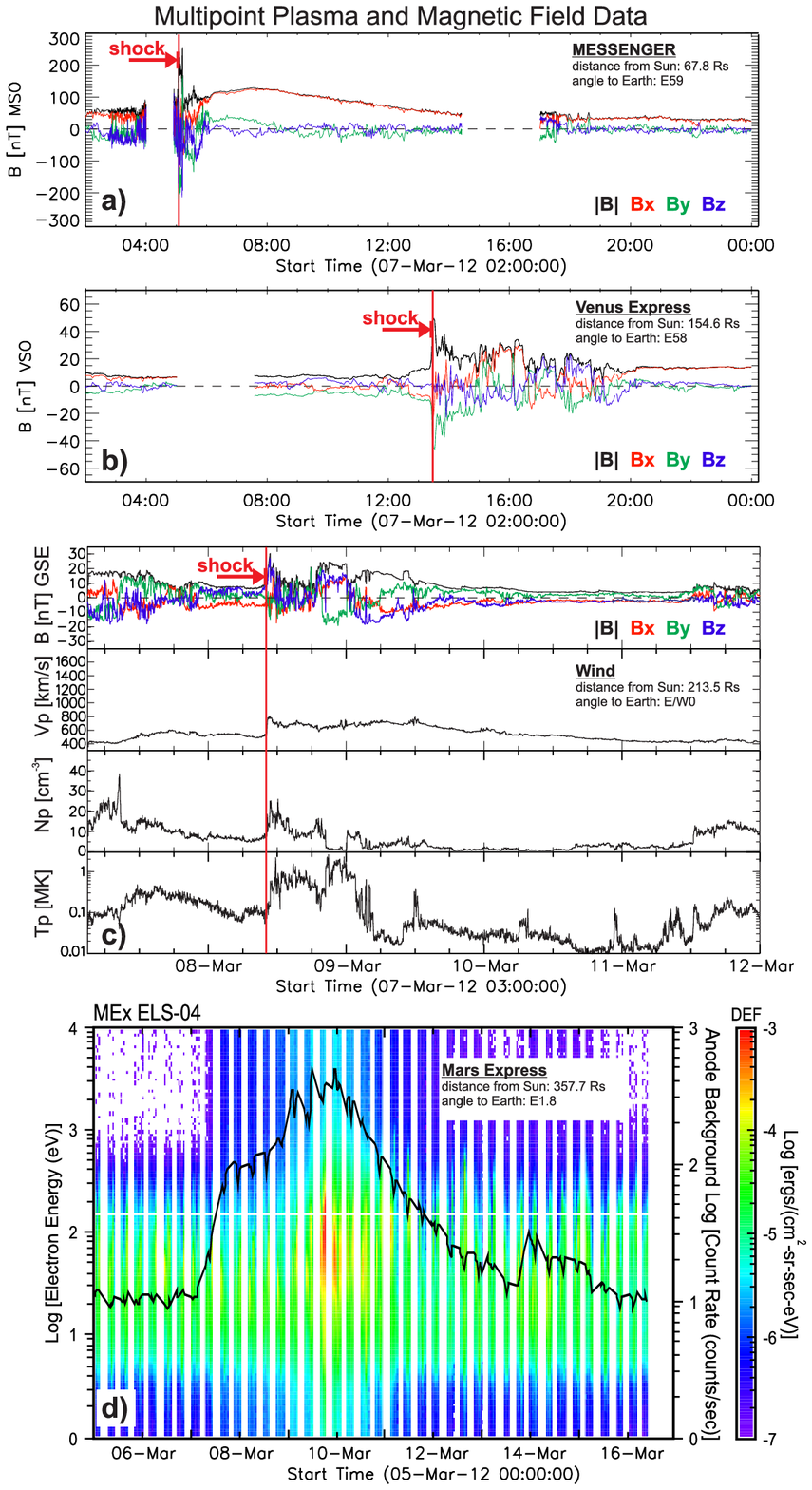}
\caption{In situ detection of the ICME at four spacecraft. a) \textit{MESSENGER} magnetic field data (MSO coordinates), b) \textit{VEX} magnetic field data (VSO coordinates), c) \textit{Wind} plasma and magnetic field data (GSE coordinates), and d)  \textit{MEX} differential electron energy flux (in color) and background count rate (black line).\label{fig2}}
\end{figure}

The ICME shock was first detected in situ by \textit{MESSENGER} at 05:00 UT (Figure \ref{fig2}a). At this time, Mercury was located at a radial distance from the Sun of $\sim 68$ $R_\odot$, $\sim59^\circ$ east of Earth. The CME shock was subsequently detected by \textit{VEX} at 13:28 UT at a heliocentric distance of $\sim 154$ $R_\odot$, $\sim 58^\circ$ east of Earth (Figure \ref{fig2}b). The shock arrival time at \textit{Wind}, which is located at the L1 point, was---according to \citet{liu13}---on 2012 March 8 at 10:19 UT (Figure \ref{fig2}c). 
The ICME arrived with a plasma speed of $\sim 800$ km s$^{-1}$ (immediately after the shock), while the average speed in the sheath region was $679 \pm 44$ km s$^{-1}$ \citep[][]{liu13,moe14}.
The last in situ arrival detection was at \textit{MEX}. Examining the flux of the 200 eV electrons, the ICME shock seemed to arrive at Mars between 12:13 and 16:18 UT on 2012 March 9 (Figure \ref{fig2}d). Comparing the last orbital measurement of solar wind 200 eV electron plasma flux, which occurred at 12:13 UT, with the first measurement from the next spacecraft orbit at 16:18 UT it shows an increase of over one order of magnitude. The background signature (black line) indicates penetrating radiation, produced by protons greater than 1 MeV in this case. Penetrating particle radiation is observed to encounter the spacecraft before the CME arrival and is generally an indicator of SEPs \citep[][]{fra13}.

CME4 started shortly after the onset of CME3 from the eastern edge of the same active region and was first visible in COR1-A on 2012 March 7 at 01:05 UT and in COR2-A at 01:39 UT. Within the HI FoV it is not possible to distinguish between those two events.

\begin{table*}
\begin{center}
\caption{First white-light observations and in situ arrival times at \textit{MESSENGER}, \textit{VEX}, \textit{Wind} and \textit{MEX} of CME3 and the three associated CMEs.}
\begin{tabular}{lcccccc} 
\toprule
Event & COR1-A & COR1-B & \textit{MESSENGER} & \textit{VEX} & \textit{Wind} & \textit{MEX} \\ 
\toprule 
                                    &   2012/03/05   & 2012/03/05 &      2012/03/05    & 2012/03/06        & -- & --\\
\multirow{-2}{*}{CME1}&    03:10 UT   &    02:46 UT   & 12:30 UT               & 08:30 UT             & --  & --\\
\midrule
                                   &      2012/03/06   &    2012/03/06   &             --                 &               --             & -- &  --\\
\multirow{-2}{*}{CME2}&      03:25 UT       & 02:36 UT &               --              &           --                 &  --&   -- \\
\midrule
                                   &        2012/03/07      & 2012/03/07 &      2012/03/07    & 2012/03/07        & 2012/03/08  & 2012/03/09  \\
\multirow{-2}{*}{\bf{CME3}}&        00:25 UT        &    00:16 UT   & 05:00 UT               & 13:28 UT             &  10:19 UT& 12:13--16:18 UT \\
\midrule
                                   &      2012/03/07        &  2012/03/07  &   --      &    --    & -- &  --\\
\multirow{-2}{*}{CME4}     &       01:15 UT      & 01:16 UT  &    --     &    --    &--  &  --\\
\bottomrule
\end{tabular}
\label{tbl-1}
\end{center}
\end{table*}

\section{Methods}

We are interested in the kinematics of CME3 and how its propagation and overall structure was influenced by the solar wind and preceding CMEs. CME3 was imaged from \textit{STEREO-A} as well as from \textit{STEREO-B} and we consider both observations independently. We apply a single-spacecraft method, based on the CHM method, and adapt it to the SSE geometry. We constrain the kinematics derived from \textit{STEREO-A} imagery with the in situ shock arrival time at \textit{Wind} and relate them to the arrival time at \textit{MEX}. The kinematics derived from \textit{STEREO-B} imagery are constrained with the shock arrival time at \textit{VEX} and compared to the arrival time at \textit{MESSENGER}. Using the kinematic profiles obtained for the two observed segments of the CME, i.e., one propagating toward Earth/Mars and one toward Mercury/Venus, we apply the DBM to find the best fitting drag-parameter. This gives us a deeper insight into the drag conditions acting on different segments of the CME.

\subsection{Constrained Self-similar Expansion Method}

An expression for the radial distance corresponding to a specific elongation for the SSE geometry was derived by \citet[][]{dav12}. The SSE geometry assumes a circular CME shape with a half-width $\lambda$ and can be used to convert the time-elongation profile of a CME into a radial distance to reveal its kinematical profile:

\begin{equation}
R_{\rm SSE}(t) = \frac{d_{\rm o} \sin \epsilon(t)(1 + \sin \epsilon(t))}{\sin (\epsilon(t)+\phi)+\sin \lambda},
\label{eq:sse}
\end{equation}

\noindent where $R_{\rm SSE}(t)$ is the resulting radial distance from Sun-center, $d_{\rm o}$ is the Sun-observer distance,  $\epsilon(t)$ is the elongation angle of the CME front, $\phi$ is the propagation direction relative to the Sun-observer line, and $\lambda$ is the CME's angular half-width. For our study, we use $\lambda=45^\circ$. The novel CSSE method uses the additional information of the in situ arrival time to estimate the most likely CME direction. With the knowledge of the shock arrival time at a certain location in interplanetary space and the corresponding elongation of the CME shock front measured from the image at arrival time, we can constrain the kinematics by varying the direction, $\phi$, until the best match of white-light and in situ arrival time is found. A detailed description and a test of this approach, using the HM geometry, is given in \citet{rol12,rol13}.

\subsection{Drag-based Model}

The DBM assumes that the interplanetary propagation of a CME beyond 15--30 $R_\odot$ is mainly influenced by the drag-force, i.e., CMEs moving faster than the ambient solar wind are decelerated and slower CMEs are accelerated  \citep[][]{car96,vrs01,vrsgop02,car04}. The drag-acceleration, $a(t)$, is given as:

\begin{equation}
a(t) = - \gamma [v (t) - \omega (t)] \left\vert v (t) - \omega (t) \right\vert,
\label{eq:dbm}
\end{equation}

\noindent where $\gamma$ is the drag-parameter, $v(t)$ is the CME speed, and $\omega(t)$ is the speed of the ambient solar wind. $\gamma$, is mainly a function of the cross section and mass of the CME, and the density and speed of the ambient solar wind. Typical values for $\gamma$ range between $2 \times 10^{-8}$ and $2 \times 10^{-7}$ km$^{-1}$ and are smaller for massive CMEs than for less massive ones \citep[see][]{vrs12}. (DBM online tool: \url{http://oh.geof.unizg.hr/DBM/dbm.php})

\section{Results}

Comparing the kinematics derived by \cite{liu13} and \cite{dav13} with the in situ arrival time at \textit{VEX}, the CME in fact arrived between 7--14 hours earlier than derived from triangulation methods, depending on the used angular width. In our study, we treat \textit{STEREO-A} and \textit{B} observations independently and reconstruct the asymmetric global shape of the CME using the in situ arrival times at all four in situ spacecraft.

\begin{figure}
\epsscale{0.8}
\plotone{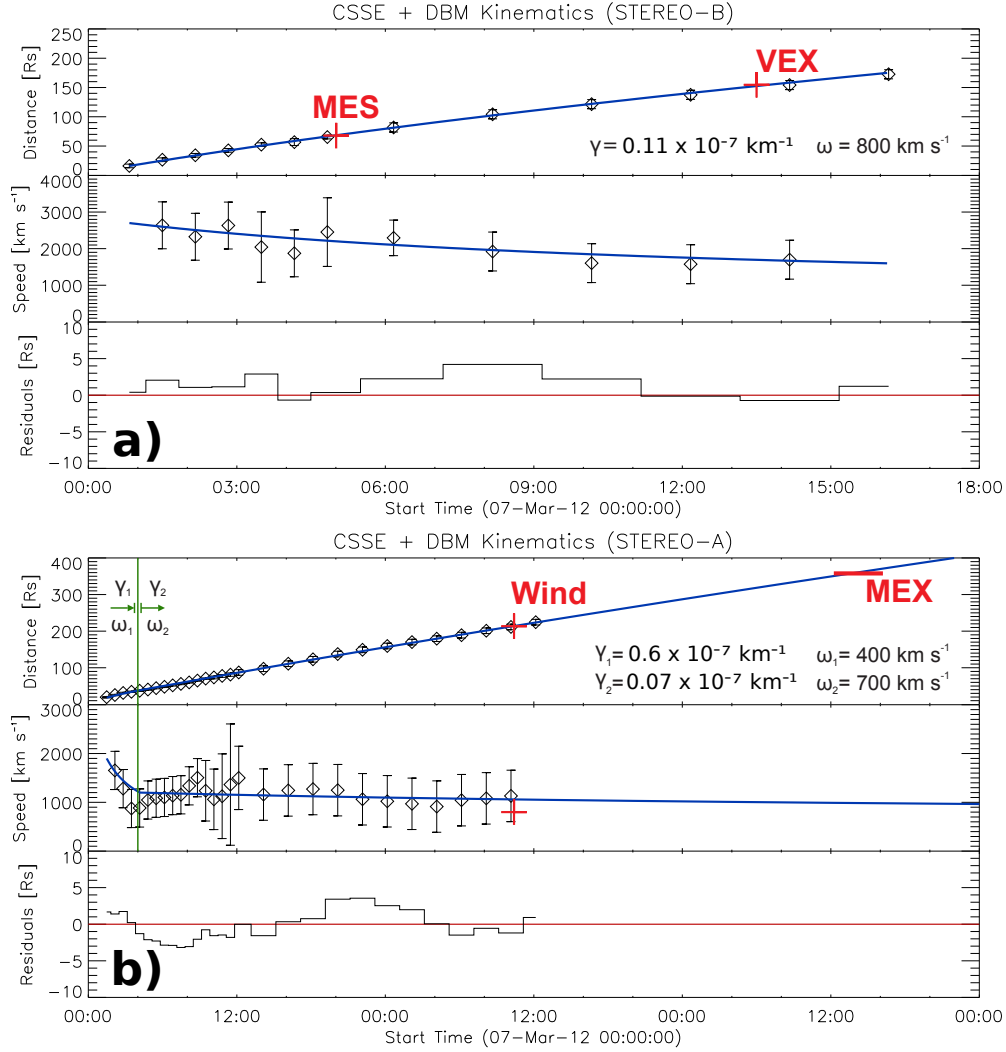}
\caption{Kinematics of CME3 derived from both \textit{STEREO} spacecraft. a) Top: CSSE time-distance profile obtained from \textit{STEREO-B}/HI observations (diamonds) and the results of the DBM (blue line). The red crosses mark the arrival times at \textit{MES} and \textit{VEX}. Middle: speed profile derived from the CSSE and DBM. Bottom: residuals between the CSSE and DBM time-distance profiles. b) Same as a) but for \textit{STEREO-A} observations. The red cross marks the shock arrival time (and speed in the middle panel) at \textit{Wind} and the red horizontal line indicates the arrival at \textit{MEX}. The green vertical line shows the time when the $\gamma$-value is changed in the DBM. \label{fig3}}
\end{figure}

Figure \ref{fig3}a shows the kinematics of CME3 derived from \textit{STEREO-B}/HI observations. The upper and middle panels show the radial distance and speed profiles, respectively, of the Mercury/Venus-directed segment of the CME front derived using the CSSE method (diamonds) and the DBM (solid line). The red crosses mark the shock arrival times at \textit{MESSENGER} and \textit{VEX}. The bottom panel displays the residuals in the radial distance between the CSSE method and the DBM. The CSSE method yields a propagation direction of E51 from Earth. By varying $\gamma$ and $\omega$ within the DBM we achieve a constant drag-parameter of $\gamma = 0.11 \times 10^{-7}$ km$^{-1}$ and a background solar wind speed of $\omega = 800$ km s$^{-1}$, which is in good agreement with the speed of CME1 of $\sim 840$ km s$^{-1}$ (derived from the travel time between Mercury and Venus). Although CME3 appears to enter the HI1-B FoV with a speed of $\sim$ 2700 km s$^{-1}$, it does not seem to undergo rapid deceleration. This could be due to the less dense solar wind left in the wake of the preceding event. Supporting evidence for this is the very faint shock front leading CME3 as seen by \textit{STEREO-B}/HI, resulting from less piled-up material in front of the flux rope. An exceptionally modest deceleration of a very fast event due to reduced drag has also been found by \cite{liu14}.

The upper panel of Figure \ref{fig3}b shows the time-distance profile of the Earth-directed segment of CME3 derived from its time-elongation profile observed by \textit{STEREO-A}/HI. In this case, the CSSE method provides a direction of motion of E11 from Earth. The results of the DBM are displayed out to the distance of Mars in order to compare it with the in situ arrival time estimated from Mars' ionospheric response. The speed derived from \textit{STEREO-A}/HI observations (middle panel) yields a different picture of this event than the results from \textit{STEREO-B}/HI observations. From the perspective of \textit{STEREO-A}, CME3 enters the HI1 FoV with a speed of $\sim 1700$ km s$^{-1}$, i.e., $\sim 1000$ km s$^{-1}$ slower than seen from \textit{STEREO-B}. The CME is still decelerating out to $\sim 40$ $R_{\odot}$ and therefore, it is necessary to adjust the drag and solar wind conditions accordingly. For $< 35$ $R_{\odot}$ we find a drag-parameter of $\gamma_{1} = 0.6 \times 10^{-7}$ km$^{-1}$ and a solar wind speed of $\omega_{1} = 400$ km s$^{-1}$. For $> 35$ $R_{\odot}$ we find the CSSE is best reproduced using the DBM with a drag-parameter of $\gamma_{2} = 0.07 \times 10^{-7}$ km$^{-1}$ and a solar wind speed of $\omega_{2} = 700$ km s$^{-1}$. This was done by manually adjusting $\gamma$ until the best agreement with the results of the CSSE method was found. The value of $\gamma$ may additionally be confirmed by determining the CME mass using, e.g., the method described in \cite{vou10}, as done by \cite{tem12} studying a CME-CME interaction process.

One explanation for the abrupt deceleration could be the preceding CME2. In Figure \ref{fig1}d, the shock front of CME2 is already at an elongation of 18$^\circ$ while the shock of CME3 is at 10$^\circ$ elongation. Nevertheless, CME2 carries a portion of dense plasma in its back, visible within HI as a bright region. This high density region may be responsible for the enhanced drag and the strong deceleration of CME3. Analyzing CME2 using the SSSE method, we found an almost constant propagation speed of $\sim 700$ km s$^{-1}$, which reflects the outcome of the DBM for the ambient solar wind speed. Another explanation for the deceleration of CME3 is given in \cite{liu13} who suggest that the rapid deceleration is caused by an energy loss due to energetic particle acceleration.

\begin{figure}
\plotone{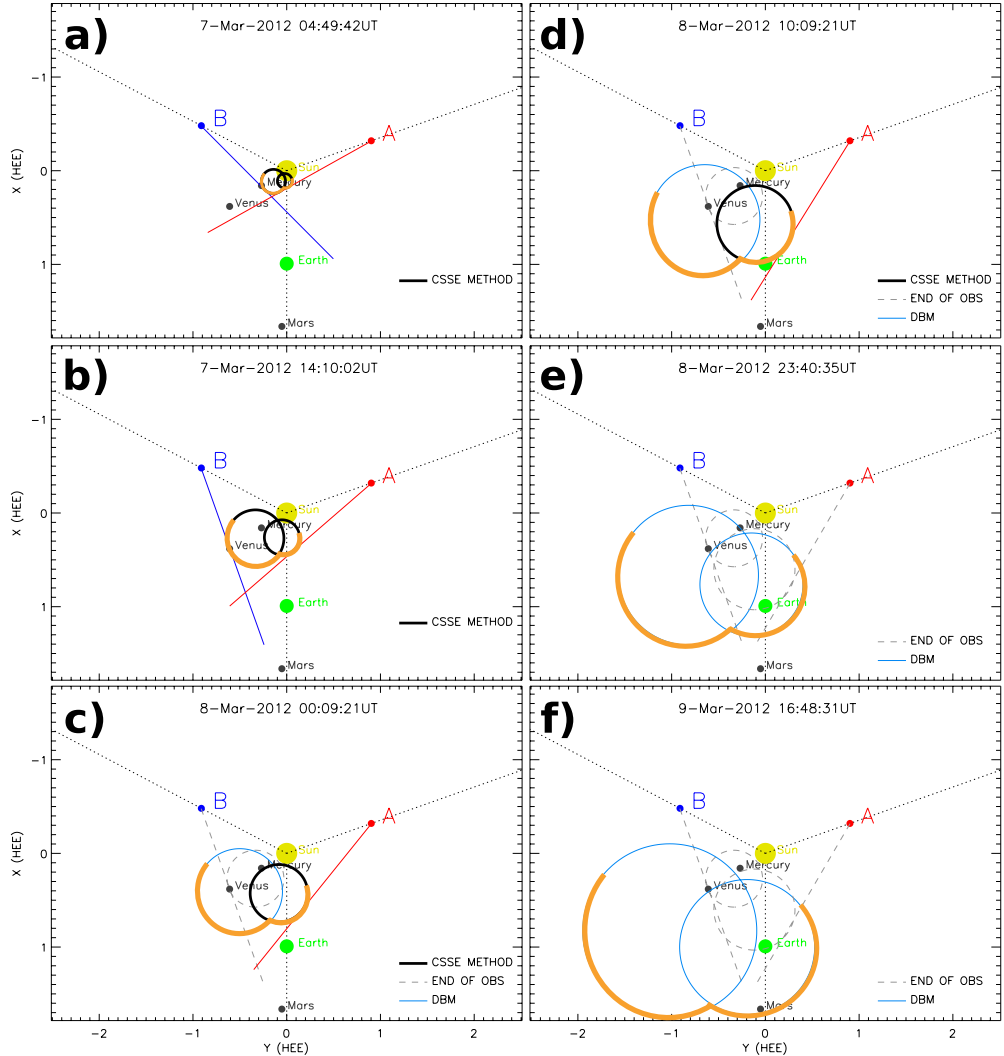}
\caption{Evolution of the shape of CME3 modeled from both \textit{STEREO} observations. The black circles show the results of the CSSE method, the gray dashed circles mark CME3 at the time of the last HI observations. The blue circles are the extrapolations of the DBM. The red and blue lines show the lines of sight from \textit{STEREO-A} and \textit{STEREO-B}, respectively. The orange area marks the reconstructed global shape of the CME front. \label{fig4}}
\end{figure}

Due to the different solar wind conditions and the resulting kinematics for the two tracks, we suggest that it is not possible to describe CME3 by only one circular front. Figure \ref{fig4} shows the evolution of the CME morphology at six different times. The blue and red lines are the lines of sight along the measured elongation angle of CME3 from \textit{STEREO-A} and \textit{STEREO-B}, respectively. The black circles are the results of the CSSE method. From the viewpoint of \textit{STEREO-B}, we are able to follow CME3 out to the location of Venus, and from the \textit{STEREO-A} vantage point we can directly observe it to the location of Earth. Beyond the  available HI observations, the blue circles represent the DBM result. The orange area displays the reconstructed global shape of the front of CME3.

\section{Discussion and Conclusions}

 The CME launched on 2012 March 7 at 00:02 UT (CME3) has the distinct advantage of observations from three different vantage points and four in situ detections, two of which almost exactly radially aligned and longitudinally separated by $\sim 67^\circ$ from the two other spacecraft. This unprecedented data set allowed us to study the evolution of CME3 in the inner heliosphere.
The eastern part of the CME, observed by \textit{STEREO-B}, shows a high speed in the sunward portion of the HI1 FoV ($\sim 2700$ km s$^{-1}$), which is slowly decreasing to $\sim 1700$ km s$^{-1}$ at a heliocentric distance of 0.7 AU. The western part of CME3, observed by \textit{STEREO-A}, entered the HI FoV with a speed of $\sim 1650$ km s$^{-1}$ and was decelerated abruptly to $\sim 1000$ km s$^{-1}$ at a heliocentric distance of $\sim 35$ $R_\odot$. Beyond this distance, this segment of the CME propagated with an almost constant speed up to 1 AU, where a shock arrival speed of $\sim 800$ km s$^{-1}$ was measured in situ by \textit{Wind}. We argue that the reason for these different speed profiles for different parts of the same CME front stems from different solar wind conditions. This idea was supported by the DBM, in which we adjusted the drag-parameter and the speed of the background solar wind to fit with the results of the CSSE method. The slow deceleration of the eastern part of the CME can be attributed to a preceding CME, i.e., the CME may have experienced less drag due to travelling through a reduced density wake of the previous event (CME1). 
At the western part, a preceding event (CME2) seemed to enhance the drag, i.e., CME3 propagated into the back of CME2 and was forced to decelerate promptly. This study emphasizes that the timing and direction of preceding CMEs may significantly alter the arrival times of CMEs at Earth. Preceding CMEs may either act (1) as obstacles, enhancing the drag and thus decelerating a CME, or (2), by creating a wake and reducing the drag, they lead to less deceleration than expected from a normal background solar wind. Sometimes---as shown for the event here---this can be true even for different parts of  the same CME.
Depending on the ambient solar wind structure, CMEs can disintegrate into distinct parts having different kinematics, yielding a deformation of their shapes and influencing the accuracy of forecasting their arrival times \citep[e.g.,][]{cro96,moe11,nie13,liu14}. Therefore, it is a goal of very high priority to investigate the evolution of the overall shape of CMEs, influenced by the ambient medium or other CMEs.



\acknowledgments{TR gratefully acknowledges the JungforscherInnenfonds of the Council of the University Graz. Support by the Austrian Science Fund (FWF): P26174-N27, V195-N16 and P21051-N16 is acknowledged by CM, TR, MT, AMV and UVA. This work has received funding from the European Commission FP7 Projects n$^\circ$ 263252 [COMESEP], n$^\circ$ 284461 [eHEROES], n$^\circ$ 606692 [HELCATS], and was supported by a FP7 Marie Curie International Outgoing Fellowship. The ASPERA-3 investigation is supported at Southwest Research Institute under NASA contract NASW-00003. We thank the \textit{STEREO SECCHI/IMPACT/PLASTIC} teams for their open data policy.}

\bibliographystyle{apj}                       

\bibliography{rollett_bib}


\end{document}